\documentclass[prl,twocolumn,showpacs,floatfix]{revtex4}
\usepackage{amssymb}
\usepackage{amsmath}
\usepackage[dvips]{graphicx}
\usepackage{epsfig}
\begin{document}
\title{Double resonances in Borromean heteronuclear triatomic systems}
\author{F. Bringas$^1$, M. T. Yamashita$^2$, T. Frederico$^3$, and Lauro Tomio$^{2,4}$}
\affiliation{
$^1$Cooperative Institute for Marine and Atmospheric
Studies, University of Miami, 4600 Rickenbacker Causeway Miami, FL
33149,  USA. \\
$^2$Instituto de F\'\i sica Te\'orica, S\~ao Paulo State University
(UNESP), 01140-070, S\~{a}o Paulo, Brazil.\\
$^3$Dep de F\'\i sica, Instituto Tecnol\'ogico de Aeron\'autica, CTA, 12.228-900, S. J. dos Campos, Brazil.\\
$^4$Instituto de F\'{\i}sica, Universidade Federal
Fluminense, 24210-346, Niter\'oi, RJ, Brazil.}
\date{\today}

\begin{abstract}
We investigate the occurrence of Borromean three-body continuum
$s$-wave resonances, in an $\alpha\alpha\beta$ system for large
negative two-body scattering lengths. The energy and width are
determined by a scaling function with arguments given by
energy ratios of the two-body virtual state subsystem energies
with the shallowest three-body bound state. The Borromean continuum
resonances emerging from Efimov
states present a peculiar behavior for trapped ultracold atoms near
a Feshbach resonance: two resonances with equal energies at
different values  of the scattering length. The corresponding
three-body recombination peaks should merge as the temperature is
raised, with one moving towards lower values of the
scattering length as the other moves to larger values.
\end{abstract}

\maketitle

The search for Efimov states was favored in recent years with the 
development of new techniques in cold-atom laboratories, such as 
laser-cooling mechanisms and the control of two-body interactions 
by Feshbach resonance techniques (for recent reports, see 
\cite{efimovnat,ferlainophys} and references therein). 
At the Innsbruck laboratory,
Efimov states are identified in scattering using ultracold gas of
Cesium atoms~\cite{kraemer} at 10 nanokelvin. And, more recently, at
the laboratory of Universit\`a di Firenze~\cite{barontini}, by using
a mixture of $^{41}$K and $^{87}$Rb it was identified the formation
of Efimov resonances in the two heteroatomic channels for KKRb and
KRbRb. The asymmetry of masses may turn more favorable the
observation of Efimov states in comparison to the case of three
identical bosons (see e.g. ~\cite{jensen,braaten,esryincao08}).

Generic systems of the type $\alpha\alpha\beta$ can be classified
according to the interaction of their two-body subsystems~\cite{jensen,raios}.
In particular, Borromean systems have been studied in \cite{borromean}.
Progress on calculating three-body continuum states and resonance decay
\cite{jensenprl} are impressive and they
give a step further in our precise understanding of these states.
Following these novel methods, applied so far in the nuclear physics
context (e.g. the decay of $0^+$ state of $^{12}$C in three
$\alpha$'s), a picture of resonances arising from Efimov states are
highly demanded in view of the tremendous advances on the
experimental setups of ultracold atoms with tunable interactions.
Some properties of these states under the change in the
interaction parameters are well known, e.g., when at least one of
the two-body subsystems is bound, an Efimov state emerges from the
two-body analytical cut by changing the two-body binding
energy~\cite{comment}. Such state, in the second energy sheet, is
given by a pole in the real axis and identified as a virtual 
state.
In the case of a Borromean three-boson system there is no two-body
cut, and the Efimov state arrives from a continuum three-body
resonance~\cite{ressonancia}. As systems of ultracold atoms
with different species offer larger opportunities for
universal physics than the equal boson case,  
it is timely to study the fate of Borromean states 
in heteronuclear systems with tunable interactions.

In the present work, we extend the analysis of
Ref.~\cite{ressonancia}, to Borromean $\alpha\alpha\beta$ systems
(negative scattering lengths) with two kind of particles. It is
shown how the Efimov states emerge from three-body $s$-wave
continuum resonances. As the absolute values of the scattering
lengths (given, respectively, by $|a_{\alpha\alpha}|$ and/or
$|a_{\alpha\beta}|$) increase, an existing three-body continuum
resonance disappears at the scattering threshold with the formation
of an Efimov bound state. This process can be replicated by further
increasing one or both two-body scattering lengths, with the
formation of more Efimov excited states. In the exact Efimov limit,
a tower with infinite excited states is produced when both,
$|a_{\alpha\alpha}|$ and $|a_{\alpha\beta}|$, turn to be infinite.

The parametric region, for which exists a three-body resonance, is
presented in the form of a scaling function having scattering
lengths measured in units of a three-body length identified with
$\left[\sqrt{m_\alpha B_3/\hbar^2}\right]^{-1}$, where $m_\alpha$ is
the mass of the $\alpha$ particle and $B_3$ is the three-body
bound-state that we use as our three-body scaling parameter. In the
following, our units will be such that $\hbar=1$ and $m_\alpha=1$, with
the mass ratio defined as
$A\equiv m_\beta/m_\alpha$.

In general terms, the quantum description of such large and weakly
bound systems is universal and can be defined by few physical
scales, despite the range and details of the pairwise interaction.
All the detailed information about the short-range force, beyond the
low-energy two-body observables, is retained in only one three-body
physical information in the limit of zero-range
interaction~\cite{jensen,braaten}. The two-body scales are defined
by the energies of the two-body virtual states, $E_{\alpha\alpha}$
and $E_{\alpha\beta}$, or by the correspondingly scattering lengths
$a_{\alpha\alpha}\simeq -1/\sqrt{|E_{\alpha\alpha}|}$ and
$a_{\alpha\beta}\simeq -1/\sqrt{|E_{\alpha\beta}|}$. Such two-body
scales are given in units of a three-body length scale
$1/\sqrt{B_3}$, with $B_3$ defined as the binding energy of the
shallowest three-body state. In the present case, the scaling
function for the continuum resonance energy $E_3$ can be written as
\begin{equation}
E_{3} = B_{3}~{\cal E}\left({a_{\alpha\alpha}} \sqrt{{B_{3}}},
{a_{\alpha\beta}}\sqrt{{B_{3}}};A\right). \label{oe}
\end{equation}
The corresponding critical boundary, given by ${\cal
E}\left({a_{\alpha\alpha}} \sqrt{{B_{3}}},
{a_{\alpha\beta}}\sqrt{{B_{3}}};A\right)=0 , $ in the parametric
plane $\left(\sqrt{|E_{\alpha\alpha}|/B_{3}},
\sqrt{|E_{\alpha\beta}|/B_{3}}\right)$, was numerically verified in
Ref.~\cite{AmPRC97}, where all possibilities for bound and virtual
two-body states were considered. We note that the calculations done
throughout this work were performed by considering only two-body
virtual states (negative scattering lengths). Once the critical
boundary is crossed in the direction of infinity negative scattering
lengths, an existing continuum three-body resonance will move to an
Efimov bound state. Such continuum three-body state was indeed
observed in the particular case of identical bosons, through the
resonant behavior of three-body recombination rate in an ultracold
trapped gas of Caesium~\cite{kraemer}.

The presence of Efimov resonances has been also reported in Ref.~\cite{barontini} 
in an ultracold mixture of $^{41}$K and $^{87}$Rb gases, through the three-body 
collision of KKRb and RbRbK atoms. In these cases, two resonantly interacting 
pairs for positive or negative scattering lengths are sufficient to allow
Efimov states.
 Such experiment supports the prediction made
 in Ref.~\cite{AmPRC97}, which is resumed in a critical boundary
 in the parametric plane defined by
 $\left(\pm\sqrt{|E_{\alpha\alpha}|/B_{3}},
 \pm\sqrt{|E_{\alpha\beta}|/B_{3}}\right),$
 for the existence of at least one Efimov bound state.
 Three-body heteronuclear mixtures with resonant interspecies were
also been studied in \cite{hetero} using zero-range interaction,
where the critical conditions for formation of Efimov states at
threshold are discussed. The case presented in this work of
continuum resonances were left open.

In order to calculate the scaling function (\ref{oe}), we use
subtracted Faddeev equations with zero-range interactions for a three-body system composed by
two identical bosons, $\alpha$, and a third one, $\beta$. The
following equations, presented for continuum resonant states, were
already written in a similar form in the context of halo nuclei
systems~\cite{raios,comment}. After partial-wave projection, the
$s-$wave coupled integral equations of the Faddeev
spectator functions $\chi_{\alpha\alpha}$ and $\chi_{\alpha\beta}$,
can be written as {\small
\begin{eqnarray}
\chi_{\alpha\alpha}(q)\hspace{-0.2cm}&=&\hspace{-0.2cm}4\pi\,\tau_{\alpha\alpha}
(q;E_3)\int_0^\infty\hspace{-0.2cm} p^2 dp\int_{-1}^1\hspace{-0.2cm}dz
G_1(q,p,z;E_3) \chi_{\alpha\beta}(p),
\nonumber\\
\label{Xab}
\chi_{\alpha\beta}(q)\hspace{-0.2cm}&=&\hspace{-0.2cm}2\pi\,
\tau_{\alpha\beta}(q;E_3) \int_0^\infty p^2 dp
\int_{-1}^1 dz \times \\ \nonumber
&\times&
\left[G_1(p,q,z;E_3)\chi_{\alpha\alpha}(p)+G_2(q,p,z;E_3)\chi_{\alpha\beta}(p)\right],
\end{eqnarray}
} where the two-body amplitudes, $\tau_{\alpha\alpha}$ and $\tau_{\alpha\beta}$, and the subtracted Green functions, $G_1$ and
$G_2$, are defined in Ref.~\cite{comment}. The subtraction energy is of no physical importance and it is let to $\infty$.

\begin{figure}[hbt]
\vspace{-1.5cm}
\centerline{\epsfig{figure=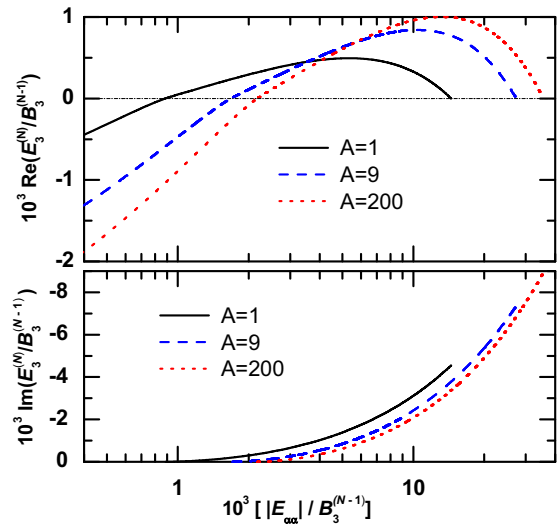,width=8cm}}
\vspace{-1.7cm} \caption[dummy0]{Real (upper frame) and imaginary
(lower frame) parts of the three-body complex energy as a function
of the two-body virtual state energy with
$E_{\alpha\alpha}=E_{\alpha\beta}$. The results, obtained for $N=1$,
are given for mass ratios $A=$1, 9, 200.} 
\vspace{-.5cm}
\label{fig1} \end{figure}

The resonance energies are calculated from the coupled eqs. (\ref{Xab})
using a contour deformation method in the complex momentum
plane~\cite{complex}. The momentum variable as
$p$ appears as a function of the deformation
angle $\theta$ and written as $p\equiv |p| e^{-{\rm i}\theta}$,
with $0\le\theta<\pi/4$. The second energy sheet is revealed when
the momentum integration path is deformed by a contour along the real axis to the complex plane with
a fixed rotation angle chosen to place the contour path far from the
scattering singularities. That exposes the complex resonance
energy pole of the scattering matrix. For large enough $\theta$, the
solution of the coupled eqs. (\ref{Xab}) in the complex energy plane
is found for $\tan(2\theta)>-Im(E_3)/Re(E_3)$, where $Re(E_3)$
represents the resonance energy and $Im(E_3)$ its half width.

In the limit when the two-body energies are equal to zero, an
infinite number of Efimov states emerges from the solution
of the coupled eqs. (\ref{Xab}). At this point, by varying the
two-body energies, a given energy $E_3$ will have zero imaginary
part, becoming purely real and negative. The $N^{th}$
Efimov bound state is defined by $E_3^{(N)}\equiv -B_3^{(N)}$ with
$N=0$ indicating the ground state. When the $N^{th}$ Efimov bound state
is moved to a resonance  its complex energy will be denoted by $E_3^{(N)}$.

\begin{figure}[thb]
\centerline{\epsfig{figure=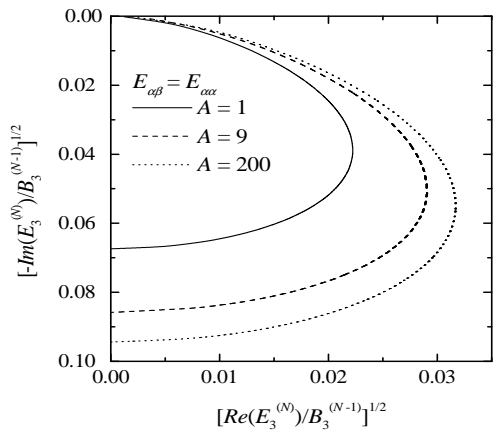,width=7.0cm,height=6cm}}\vspace{-0.5cm}
\caption[dummy0]{Given the $N-$th three-body resonant energy in
units of the $(N-1)-$th bound state energy $B_3^{(N-1)}$, it is
plotted $\sqrt{-Im(E_3^{(N)})/B_3^{(N-1)}}$ as a function of
$\sqrt{Re(E_3^{(N)})/B_3^{(N-1)}}$, for $N=1$,
$E_{\alpha\alpha}=E_{\alpha\beta}$ and mass-ratios $A=$ 1, 9 and
200. } \label{fig2}\vspace{-.5cm}
\end{figure}

The solution of the homogeneous coupled eqs. (\ref{Xab}) in the
complex rotated contour for complex energies gives the position and
width of the continuum three-body resonances, allowing to construct
numerically the scaling function (\ref{oe}). In practice, we note
that one can approach quite fast the scaling limit~\cite{FrPRA99}
or the limit cycle~\cite{mohr} for $N\to\infty$, such that we will
present results for $N=1$. In two frames, we show in Fig.~\ref{fig1}
the real and imaginary parts of the complex energy $E_3^{(N)}$, as
functions of the two-body binding energy $E_{\alpha\alpha}$, for the
case that $E_{\alpha\alpha}=E_{\alpha\beta}$. All such energy
quantities are dimensionless, given in units of the bound-state
energy $B_3^{(N-1)}$. So, in Fig.~\ref{fig1}(a), we have
$Re(E_3^{(N)})/B_3^{(N-1)}$ versus $E_{\alpha\alpha}/B_3^{(N-1)}$;
and, in Fig.~\ref{fig1}(b), $Im(E_3^{(N)})/B_3^{(N-1)}$ versus
$E_{\alpha\alpha}/B_3^{(N-1)}$.
In Fig.~\ref{fig1}(a), in the positive part of the plot we have the
$N-$th resonance energy given by $Re(E_3^{(N)})/B_3^{(N-1)}$. When
the ratio $E_{\alpha\alpha}/B_3^{(N-1)}$ is decreased, this
resonance turns into a bound state and the negative part of the plot
gives the ratio of the energies of two consecutive states
($B_3^{(N)}$) and ($B_3^{(N-1)}$) three-body bound state for each
system. This transition from a resonant to a bound state follows the
same general behavior as verified in the case of three identical
bosons~\cite{ressonancia}. For $A=1$, $9$ and $200$, the values of $E_{\alpha\alpha}/B_3^{(N-1)}$
at which we have transitions from resonances to bound-states are,
respectively, given by 0.876$\times 10^{-3}$, 1.67$\times 10^{-3}$
and 2.15$\times 10^{-3}$. These values correspond to the ones given
the position of the boundary curve in the parametric space
$\left(-\sqrt{|E_{\alpha\alpha}|/B_{3}},-\sqrt{|E_{\alpha\beta}|/B_{3}}\right)$
(for the case  $E_{\alpha\alpha}=E_{\alpha\beta}$)~\cite{AmPRC97}
that separates the excited bound Efimov state from the continuum
resonance.

In Fig.~\ref{fig1}, our plots are determined only for $N=1$,
considering that results for higher $N$'s should coincide with the
ones obtained for $N=1$.
The energy value of the resonance (real part) grows, as the virtual
two-body energies increases (by reducing the absolute values of the
scattering lengths), up to a maximum and then decreases, while the
imaginary part always increases. This makes the resonance to dive
deeper in the second energy sheet,  as it is clearly shown in
Fig.~(\ref{fig2}).

The results shown in Fig.~\ref{fig1} suggest that in an ultracold 
mixture of heteronuclear atoms the
increase of temperature moves a Borromean Efimov continuum resonance
towards smaller absolute values of the scattering length (larger
two-body virtual energies), while the width [Imaginary part of
$E_3^{(N)}$] increases. Indeed,  an experiment with ultracold
Caesium atoms~\cite{kraemer} indicated that the recombination peak
due to the continuum triatomic resonance moves towards smaller
values of $|a|$ as temperature is raised, therefore such resonance
is identified with a state that just arises from an Efimov bound one
that dives into the three-boson continuum. Such behavior is a clear
consequence of the path followed by the real part of the resonance
energy that increases as $|a|$ is decreased (see Fig.~\ref{fig1} 
and ref.~\cite{pla363}).

The maximum value reached by the real part of the three-body energy, the
position of the resonance, presumably creates a curious effect that one
could observe in the three-body recombination. As the position of the 
resonance returns to zero ($Re(E_3^{(N)})\to 0$), it may be possible to detect,
at a given temperature, the occurrence of two peaks in the three-body
recombination, corresponding to two values of the scattering length. The wider
peak should be at smaller values of the absolute value of the scattering length,
as one can easily verify by looking at Fig.~\ref{fig1}.

\begin{figure}[thb]
\vspace{-.3cm}
\centerline{\epsfig{figure=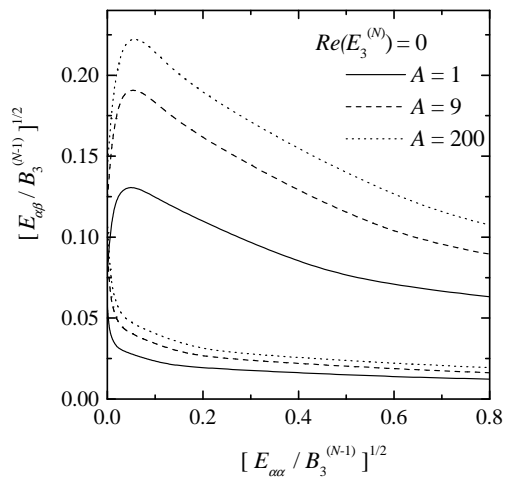,width=7.0cm}}
\caption[dummy0]{Boundaries for the existence of a resonance as a
scaling function of $\sqrt{|E_{\alpha\beta}|/B_3^{(N-1)}}$ versus
$\sqrt{|E_{\alpha\alpha}|/B_3^{(N-1)}}$. } \label{fig3}
\end{figure}

In Fig.~\ref{fig3} we show a scaling function for $E_3^{(N)}=0$ of
$\sqrt{|E_{\alpha\beta}|/B_3^{(N-1)}}$ versus
$\sqrt{|E_{\alpha\alpha}|/B_3^{(N-1)}}$, i.e., these curves are the
boundaries for the existence of a resonance, as already obtained in
Ref.\cite{AmPRC97}. In the lower section of the figure, we have
three plots carrying the points where a resonance begins to appear,
$Re(E_3^{(N)})=0$, in Fig.~\ref{fig1}; and, in  the upper section of
the figure, we have the corresponding curves carrying the points
where the origin is reached with a finite value of the width. The
boundary curves indicate that, at ultralow temperatures, a mixture
of heteronuclear gas with two species can have two three-body
recombination peaks associated with one Efimov state. As
$a_{\alpha\beta}$ is modified with $a_{\alpha\alpha}$  kept fixed, a
wider resonance appears when the upper curve in Fig.~ \ref{fig3} is
crossed, that happens for smaller values of $|a_{\alpha\beta}|$;
with a thinner one occurring for larger values. This last case
corresponds to an excited bound Efimov state diving into the
continuum. As $T$ is raised, the two peaks tend to merge, as the
maximum of the resonance  is approached.

We found two continuum resonances for Borromean systems
at different values of the scattering length and same position in
energy and different widths, for  heteronuclear and homonuclear
systems. The absolute maximum for the resonant energy (associated
with the system temperature) occurs in both cases.
As we observe in Fig.~\ref{fig1} the resonance energy attains a
maximum value and then decreases as $|a|$ is moved further towards
small values. Therefore, two resonances are located at the same
energy for different values of the negative scattering length. It is
expected that the corresponding pair of three-body recombination
peaks merges as the temperature is raised, one peak moves towards
lower values of the scattering length while the other one to larger
values. Using the numerical values of the scattering length for
homonuclear systems, one gets that  the ratio between the scattering
lengths where the real part of the resonance energy is identical
stays between 1 and 0.25 as one easily gets by inspection of 
Fig.~\ref{fig1}. Therefore, a second three-body recombination peak
appears in a region where the four-body recombination peak is present
with a scattering length ratio of 0.43 \cite{incao,ferlainophys}, posing an 
experimental challenge for separating the three- and four-body 
recombination peaks.

In conclusion, we studied the universal properties of  three-body
continuum resonances for heteronuclear three-particle system with
two species of atoms, in a Borromean configuration such that all the
two-body subsystems are unbound, i.e., each pair has a virtual state
close to the two-body scattering threshold. It is shown how a
Borromean Efimov excited bound state turns out to a resonant state
by tuning the virtual two-body subsystem energies or scattering
lengths, with all energies written in units of the next deeper
shallowest Efimov state energy. The resonance position and width for
the decay into the continuum are obtained as universal scaling
functions (limit cycle) of the dimensionless ratios of the two and
three-body scales, which are calculated numerically within a
zero-range renormalized three-body model. We discussed how the
continuum resonances for heteronuclear Borromean systems may be
observed in mixtures of ultracold gases. The fingerprint of the
observation of the scaling considered in the present work, for a
Borromean system, is that such resonances will be given by their
characteristic displacement with temperature $T$: as $T$ raises the
recombination peak, that comes from an Efimov state diving into the
continuum, should move to lower  absolute values of the scattering
lengths with an increasing width. Moreover, we found theoretically
two recombination peaks at $T=0$, with different widths and
positioned at different values of the scattering lengths. For a
fixed $\alpha\alpha$ scattering length, with the assumption that
three-body scale does not move, the wider resonance should appear
for smaller  absolute values of the $\alpha\beta$ scattering length.
As $T$ is raised the two recombination peaks tend to merge.
This physical effect may be observed in ultracold trapped
heteronuclear systems near a Feshbach resonance, where the
properties of the continuum resonances offer a rich structure to be
explored by tuning the large and negative scattering lengths. It is
an exciting possibility to be probed in  actual experiments with
mixtures of ultracold atomic gases.

We thank Funda\c c\~ao de Amparo \`a Pesquisa
do Estado de S\~ao Paulo and Conselho Nacional de Desenvolvimento
Cient\'\i fico e Tecnol\'ogico for partial financial support.


\vspace{-0.5cm}
\begin{thebibliography}{}
\bibitem{efimovnat} V. Efimov, Nature Phys. {\bf 5}, 533  (2009);
S. E. Pollack, D. Dries, and R. G. Hulet, Science {\bf 326},
1683 (2009).
\bibitem{ferlainophys}F. Ferlaino and R. Grimm, Physics {\bf 3}, 9 (2010). 
\bibitem{kraemer} T. Kr\"amer {\it et al.}, Nature {\bf 440}, 315 (2006).
\bibitem{barontini} G. Barontini {\it et al.}, Phys. Rev. Lett. {\bf 103}, 043201 (2009).
\bibitem{jensen} A.S. Jensen, K. Riisager, D.V. Fedorov, E. Garrido,
Rev. Mod. Phys. {\bf 76}, 215 (2004).
\bibitem{braaten} E. Braaten, H.-W. Hammer,  Phys. Rep. {\bf 428}, 259 (2006)
\bibitem{esryincao08} B. D. Esry, Y. Wang, and J.P. D'Incao,
Few-Body Syst. {\bf 43}, 63 (2008).
\bibitem{raios} M. T. Yamashita, L. Tomio and T. Frederico, Nucl. Phys.
A {\bf 735}, 40 (2004); 
T. Frederico, M. T. Yamashita, L. Tomio, Few-Body Syst. {\bf 45}, 215 (2009).
\bibitem{borromean} J.-M. Richard and S. Fleck, Phys. Rev. Lett. {\bf 73}, 
1464 (1994); S. Moszkowski, S. Fleck, A. Krikeb, L. Theussl, J.-M. Richard, K. Varga
Phys. Rev. A {\bf 62}, 032504 (2000).
\bibitem{comment} M. T. Yamashita, T. Frederico, L. Tomio, Phys. Rev. Lett.
{\bf 99}, 269201 (2007); Phys. Lett. B {\bf 660}, 339 (2008); Phys.
Lett. B {\bf 670}, 49 (2008).
\bibitem{ressonancia} F. Bringas, M. T. Yamashita and T. Frederico.
Phys. Rev. A {\bf 69}, 040702(R) (2004).
\bibitem{jensenprl} R. Alvarez-Rodriguez, A.S. Jensen, D.V. Fedorov, H.O.U. Fynbo,
E. Garrido, Phys. Rev. Lett. {\bf 99}, 072503 (2007);
R. Alvarez-Rodriguez, H.O.U. Fynbo, A.S. Jensen, E. Garrido,
Phys. Rev. Lett. {\bf 100}, 192501 (2008);
P. Barletta, C. Romero-Redondo, A. Kievsky, M. Viviani, E. Garrido,
Phys. Rev. Lett. {\bf 103}, 090402 (2009).
\bibitem{AmPRC97} A. E. A. Amorim, T. Frederico and L. Tomio, Phys. Rev.
C {\bf 56} R2378 (1997).
\bibitem{hetero}K. Helfrich, H.-W. Hammer, and D. S. Petrov, Phys.
Rev. A {\bf 81}, 042715 (2010).
\bibitem{complex} J. Aguilar and J.M. Combes, Commun. Math. Phys.
{\bf 22}, 269 (1971); E. Balslev and J.M. Combes, Commun. Math.
Phys. {\bf 22}, 280 (1971).
\bibitem{FrPRA99} T. Frederico, L. Tomio, A. Delfino, and A. E. A. Amorim,
Phys. Rev. A {\bf 60}, R9 (1999);
A. Delfino, T. Frederico, and L. Tomio, Few-Body Syst. {\bf 28},
259 (2000).
\bibitem{mohr} 
S. Albeverio, R. Hoegh-Krohn and T.S. Wu, Phys. Lett. A {\bf 83}, 105 (1981).
R. F. Mohr, R. J. Furnstahl, H.-W. Hammer, R. J. Perry, K. G.
Wilson, Annals of Phys. {\bf 321}, 225 (2006).
\bibitem{pla363} M.T. Yamashita, T. Frederico, and L. Tomio,
Phys. Lett. A{\bf 363} (2007) 468.
\bibitem{incao} J. von Stecher, J. P. D'Incao, C. H. Greene,  Nature Phys. {\bf 5},
417 (2009).
\end{thebibliography}
\end{document}